\documentclass{article}

\usepackage{arxiv}

\usepackage[utf8]{inputenc} 
\usepackage[T1]{fontenc}    
\usepackage{hyperref}       
\usepackage{url}            
\usepackage{booktabs}       
\usepackage{amsfonts}       
\usepackage{nicefrac}       
\usepackage{microtype}      
\usepackage{lipsum}		
\usepackage{graphicx}

\usepackage{csquotes}

\usepackage{graphicx}
\usepackage{subcaption}
\usepackage{todonotes}
\usepackage[mathscr]{euscript}  
\usepackage{amsfonts}           

\title{Towards Tool-Support for Interactive-Machine Learning Applications in the Android Ecosystem}

\author{ \\Muhammad Mehran Sunny$^{1,3}$, Moritz Berghofer $^{2,3}$,  Ilhan Aslan$^{3}$\\
$^1$ Otto von Guericke University of Magdeburg\\
$^2$ University of Augsburg\\
$^3$ Huawei Technologies, Munich Research Center\\
Germany\\
$^1$ \texttt{muhammad.sunny@st.ovgu.de} \\ 
$^2$ \texttt{moritz.berghofer@student.uni-augsburg.de} \\
$^3$ \texttt{name.lastname@huawei.com} \\
}

\begin{document}
\maketitle

\begin{abstract}
Consumer applications are becoming increasingly smarter and most of them have to run on device ecosystems. Potential benefits are for example enabling cross-device interaction and seamless user experiences. 
Essential for today's smart solutions with high performance are machine learning models. However, these models are often developed separately by AI engineers for one specific device and do not consider the challenges and potentials associated with a device ecosystem in which their models have to run. We believe that there is a need for tool-support for AI engineers to address the challenges of implementing, testing, and deploying machine learning models for a next generation of smart interactive consumer applications.
This paper presents preliminary results of a series of inquiries, including interviews with AI engineers and experiments for an interactive machine learning use case with a Smartwatch and Smartphone.  
We identified the themes through interviews and hands-on experience working on our use case and proposed features, such as data collection from sensors and easy testing of the resources consumption of running pre-processing code on the target device, which will serve as tool-support for AI engineers.		
	
\end{abstract}

\section{Introduction}
Progress in technology has been inspiring new trends and interaction paradigms in the field of human-computer interaction. However, in early stages of new trends, designing and exploring user experiences is cumbersome and thus fellow researchers offer tool-support and development kits to help anyone interested to easily contribute to the exploration of new interaction paradigms. One example from the past are tangible user interfaces for which it was very hard to develop prototypes without any expertise in electrical engineering. Today, a multitude of tool-kits, such as Arduino are on the market which enable users from all walks of life to build and explore tangible interfaces easily.   

One of today's important trends is augmenting arguably any application with machine learning (ML) models, which have proven to outperform many hand-coded techniques for recognition tasks, such as recognizing objects in an image \cite{6821513}. 
Previous research has demonstrated based on prototypical systems the benefits of mobile multi-display and multi-modal assistants for tourist guides and navigation systems \cite{bum2004, compass2008}. Today, software components for intelligent application behavior are being integrated increasingly in mobile consumer applications to enable user and context aware products and provide personalized user experience. Recent ML frameworks such as TensorFlow Lite and new Neural Processing Units on mobile devices already enable easier deployment of complex ML models on mobile devices. 

However, there are specific challenges when ML-based features need to be integrated into consumer products, such as using edge machine learning, which can address connectivity, latency, scalability, privacy and security challenges. Furthermore, for consumer applications, user-centered ML techniques need to be implemented, such as interactive machine learning (IML), in which machine learning models are modified based on user input even after deployment, e.g., to adapt ML models to individual behaviors \cite{10.1145/3196709.3196729}. Last but not least, model updates in IML are more rapid, focused, and incremental \cite{Amershi_Cakmak_Knox_Kulesza_2014}. 
We believe future consumer applications will have to run on a device ecosystem. Today, for AI engineers (and in more general smart application design teams) targeting a device ecosystem has many challenges, including the collection of exemplary sensor data and rapid prototyping of functional prototypes. We believe that there is a need for tool-support for rapid prototyping of IML applications, which can be addressed. This tool support shall provide a testbed for AI engineers and address AI engineers' pain points related to the mobile ecosystem, such as data collection, experience sampling techniques, device-specific UIs and resource-sensitive data management for in-built device sensors \cite{10.1145/2971648.2971711,10.1145/3196709.3196729}.

Towards understanding today's challenges and potentials that especially AI engineers face, we conducted a series of inquiries. In the following we summarize the preliminary results, which are based on a literature study, interviews with AI engineers, and our own experiences from a small scale smartwatch-based gesture recognition project.  

\section{Challenges of Machine Learning for Interactive Consumer Application}

\begin{figure}[]
   \centering
  \includegraphics[width= 0.9\textwidth]{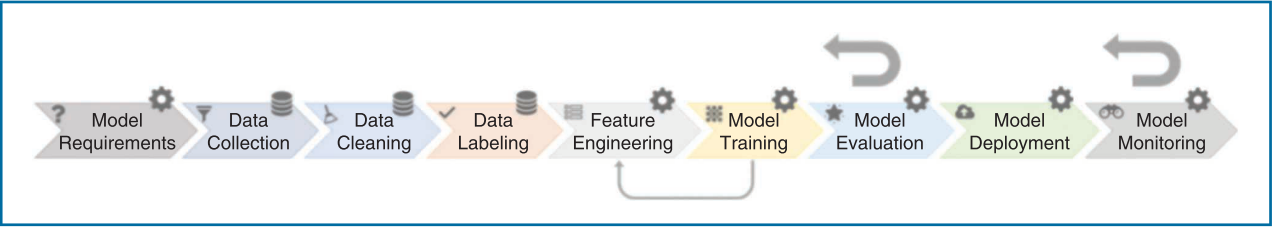}
   \caption{The nine-step process used at Microsoft to build and deploy machine-learning applications.}
    \label{fig:microsoft_ML_process}
 \end{figure}

\begin{figure}[]
   \centering
  \includegraphics[width= 0.9\textwidth]{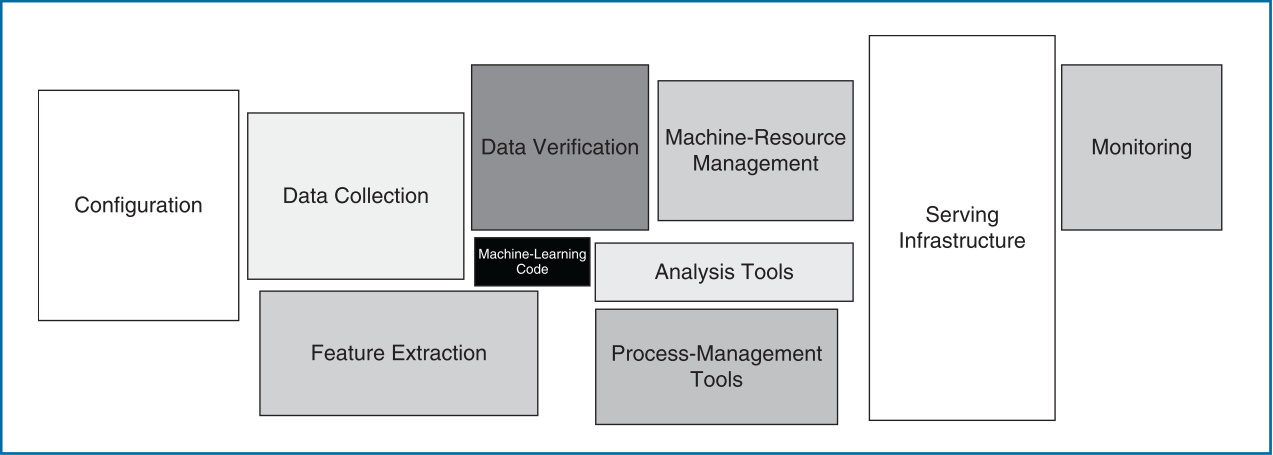}
   \caption{Lines of code in Google’s AI-related software. Other than the black box the diagram shows support software.}
   \label{fig:gooogle_AI}
\end{figure}

Various challenges have been considered in previous work to facilitate the development of ML applications on consumer devices \cite{BaierJoehrenSeebacher2019_1000095028}. In Figure \ref{fig:microsoft_ML_process}, Saleema Amershi and her colleagues at Microsoft describe their industrial-AI work as a nine-step pipeline \cite{10.1109/MS.2019.2954841}. Data modeling, cleaning, and pre-processing require extensive software development support on consumer devices because consistency, accuracy, and completeness of data are essential for ML models. Therefore, data management principles and practices need to be adopted for tool-support design \cite{8906736}. Model deployment and monitoring on consumer devices also require support from software engineers. Figure \ref{fig:gooogle_AI} shows the software engineering support for Google’s AI-related software \cite{10.1109/MS.2019.2954841}.
According to StackOverflow, the most common software engineering challenges in the AI project process are application crash, model deployment, software compatibility, installation, and performance \cite{8987482}.

\subsection{Needs of Today's IML Engineers}
IML and its tool-support for AI engineers is a relatively new research field. 
Tool-support is essential to implement and deploy IML solutions on consumer devices as AI engineers often lack software development skills for consumer devices. Currently, the majority of available tool-support is for the typical machine-learning process. The IML process demands more software development challenges as users interact directly by device-specific UIs to drive the model behavior on consumer devices. 
For tool-support design, we first identify relevant challenges with a structured literature analysis of previous research work. Subsequently, we conduct an interview study with four AI engineers across different domains, perform a qualitative analysis, and organize the challenges into five categories. Eventually, we evaluate the results from both the literature and interview with thematic analysis. We identify key issues AI engineers face in the IML process. We also perform multiple computational and communicational analyses across different android devices, i.e., smartphones and smartwatches, for different IML processes and produce benchmarks to better understand these devices' capabilities.

\subsubsection{Literature Study}
The literature study is divided according to the five steps of the AI project lifecycle: \textbf{(i) Data collection} can be very expensive. Once the data is collected, the AI engineer has to verify if the data is sufficient for the use case \cite{8906736}. Analyzing the collected data is the most critical step as AI engineers do not want their system to make false predictions \cite{URLDataCollection}. \textbf{(ii) Data pre-processing} needs to be done, for example to filter missing values or extract and rearrange the data to the model's needs \cite{10.1109/MS.2019.2954841}. Computation and power resources management are essential factors on consumer devices with limited resources \cite{Arpteg_2018}. \textbf{ (iii) ML model deployment} on consumer devices requires a good understanding of the device OS and available frameworks to facilitate this task \cite{BaierJoehrenSeebacher2019_1000095028, 8987482}. A good mix of software engineering and ML knowledge is therefore a prerequisite. \textbf{(iv) User interface design} is an essential component of IML to enable interaction with the users \cite{URLSoftwareDeployment}. A good interface design can provide a better understanding and transparency of the ML model on consumer devices for the user. Tool support is expected as UI on consumer devices developed by the software development team. \textbf{(v) Monitoring and Observability:} A good performance in the lab does not automatically translate into real-world scenarios. Therefore, monitoring and observability are essential for developing the right solution. ML model accuracy across multiple devices is very challenging and desirable at the same time. An excellent understanding of AI and SE skills is required to achieve this goal \cite{10.1109/CESSER-IP.2019.00009}.

\subsection{Interview Study}
The interview study has been divided into five steps to understand the requirements of tool-support for the AI process lifecycle on consumer devices:  \textbf{(i) Data collection:} Standard sensors used in data collection are IMU sensors (Accelerometer and Gyroscope), Cameras, microphones, and GPS. AI engineers face difficulties accessing these sensors and inconsistent sample rates on consumer devices, with very limited computational, power, and communicational resources, especially in multimodal data collection. There is a lack of available libraries or frameworks to facilitate this process. Tool-support is essential to collect data from multiple sensors and UIs on consumer devices and access the data in an easily readable format for analysis. \textbf{(ii) Data pre-processing:} AI engineers are often unaware of consumer device capabilities, available resources, and the development environment. Data pre-processing requires heavy computation, and it is challenging to manage resources. AI engineers have no control over available libraries and support for their research needs. The tool-support design requirements are data synchronization, resource optimization, and an easily configurable UI for data cleaning, pre-processing, and storage. \textbf{ (iii) ML model deployment:} Model size and compatibility with the consumer devices are challenging as they can increase the application size and computation resources requirements. TensorFlow Lite support has made the model deployment process easy on consumer devices, but the model's conversion from other frameworks to TensorFlow Lite is complicated. AI engineers require a tool-support with guidelines to perform model conversion and deployment on edge devices. Tool-support should also help in deploying updated versions of the model in already existing solutions. \textbf{(iv) User interface design} to interact with the user is an essential part of the IML process. UI on consumer devices should be consistent, user-friendly, engaging, and provide a good understanding of the model to the end-user. UI optimization is required to save on-device resources. The tool-support can help in an easy-to-configured dynamic user interface to collect data and labels from users across multiple devices. \textbf{(v) Monitoring and Observability:} Observing the model accuracy and performance after deployment across multiple devices is a particular interest for AI engineers. In IML, we cannot directly rate the model's performance as it is subjective and driven by end-user behavior. However, tool-support can take user feedback or satisfaction by some means, it can guide the AI engineers to improve model accuracy across multiple devices by some standardization. 
We have provided a concise summary of the interviews performed. The preliminary themes we have identified are lack of multi-devices expertise, scientific experimental challenges, user interaction, and on-device ML support.

\section{Conclusion}
Today's paradigmatic ecosystems consist of at least a smartwatch and a smartphone. But even this seemingly simple ecosystem will be associated with many challenges for consumer application developers who aim to provide smart and seamless experiences. We believe that machine learning will be of core value for such smart and seamless experiences and that there is a need for tool-support for interactive machine learning applications. Enabling AI engineers and more generally designers with AI abilities to quickly create functional prototypes and explore ideas will significantly accelerate progress in this domain. This paper presented first results and insights towards our ongoing research in the creation of tool-support for IML engineers. We hope to inspire fellow researchers and beyond that ignite a fruitful and critical discussion on this topic.

\bibliographystyle{unsrt}
\bibliography{Literature}

\end{document}